\documentstyle[10pt,epsf,epsfig,dp_delphititle,oldlfont,amssymb]{dp_delphi}
\makeindex
\def\DpPaperGroup{PH-EP}
\def\DpPaperRef{2007-017}
\def\DpDate{25 May 2007}
\def\DpAuthors{DELPHI Collaboration}
\def\DpTitle{{ Search for Pentaquarks in the Hadronic Decays of
the Z Boson with the DELPHI Detector at LEP }}
\def\DpSubmit{(Accepted by Phys. Lett. B)}
\def\DpComment{ }
\def\DpEMail{ }

\def\jpsi{\hbox{$J{\kern-0.24em}/{\kern-0.14em}\psi$}}
\def\ev#1#2{\hbox{#1e{\kern-0.10em}V{\kern-0.30em}/{\kern-0.14em}$#2$}}

\def\bqt#1#2\eqt{\begin{equation}\label{#1}%
        {#2}\end{equation}\noindent}
\def\bln#1#2\eln{\begin{equation}\label{#1}%
            \eqalign{#2}\end{equation}\noindent}
\def\brlist{\begin{thebibliography}{99}}

\def\erlist{\end{thebibliography}}

\begin{document}
\makeatletter
\makeatother

\begin{titlepage}
\pagenumbering{roman}

\CERNpreprint{\DpPaperGroup}{\DpPaperRef}   
\date{{\small\DpDate}}              
\title{\DpTitle}                
\address{\DpAuthors}                

\begin{shortabs}                
\noindent
The quark model does not exclude states composed of  more than three
quarks, like pentaquark systems.  Controversial 
evidence for such states has been published in the last years, in particular: for a strange 
pentaquark $\Theta(1540)^+$; for a double-strange state, the $\Xi(1862)^{--}$, subsequently called
$\Phi(1860)^{--}$; and for a charmed state, the $\Theta_c(3100)^{0}$. If 
confirmed, a full pentaquark family might exist; such pentaquark states could 
be produced in $e^+e^-$ annihilations near the Z energy. 
In this paper a search for pentaquarks is described using the DELPHI 
detector at LEP, characterized by powerful particle
identification
sub-systems crucial in the separation of the signal from the
background
for these states. At 95\% CL, upper limits are set on the 
production rates $\langle N \rangle$ of such particles and their charge-conjugate state per 
Z decay:
\begin{eqnarray*}
\langle N_{\Theta^{+}} \rangle  \times Br(\Theta^+ \rightarrow pK^0_S) & < & 5.1 \times 10^{-4} \\
\langle N_{\Theta^{++}} \rangle & < & 1.6 \times 10^{-3} \\
\langle N_{\Phi(1860)^{--}} \rangle \times Br(\Phi(1860)^{--} \rightarrow \Xi^- \pi^-) & < & 2.9 \times 10^{-4} \\
\langle N_{\Theta_c(3100)^{0}} \rangle \times Br(\Theta_c(3100)^0 \rightarrow D^{*+}\bar p) & < & 8.8 \times 10^{-4}  \; .
\end{eqnarray*}
\end{shortabs}

\vfill

\begin{center}
\DpSubmit \ \\      
\DpComment \ \\
\DpEMail \ \\
\end{center}

\vfill
\clearpage

\headsep 10.0pt

\addtolength{\textheight}{10mm}
\addtolength{\footskip}{-5mm}
\begingroup
%
\newcommand{\DpName}[2]{\hbox{#1$^{\ref{#2}}$},\hfill}
\newcommand{\DpNameTwo}[3]{\hbox{#1$^{\ref{#2},\ref{#3}}$},\hfill}
\newcommand{\DpNameThree}[4]{\hbox{#1$^{\ref{#2},\ref{#3},\ref{#4}}$},\hfill}
\newskip\Bigfill \Bigfill = 0pt plus 1000fill
\newcommand{\DpNameLast}[2]{\hbox{#1$^{\ref{#2}}$}\hspace{\Bigfill}}

%
\footnotesize
\noindent
\DpName{J.Abdallah}{LPNHE}
\DpName{P.Abreu}{LIP}
\DpName{W.Adam}{VIENNA}
\DpName{P.Adzic}{DEMOKRITOS}
\DpName{T.Albrecht}{KARLSRUHE}
\DpName{R.Alemany-Fernandez}{CERN}
\DpName{T.Allmendinger}{KARLSRUHE}
\DpName{P.P.Allport}{LIVERPOOL}
\DpName{U.Amaldi}{MILANO2}
\DpName{N.Amapane}{TORINO}
% ...
\normalsize
\endgroup
\newpage

\titlefoot{Department of Physics and Astronomy, Iowa State
     University, Ames IA 50011-3160, USA
    \label{AMES}}
\titlefoot{IIHE, ULB-VUB,
     Pleinlaan 2, B-1050 Brussels, Belgium
    \label{BRUSSELS}}
\titlefoot{Physics Laboratory, University of Athens, Solonos Str.
     104, GR-10680 Athens, Greece
    \label{ATHENS}}
\titlefoot{Department of Physics, University of Bergen,
     All\'egaten 55, NO-5007 Bergen, Norway
    \label{BERGEN}}
\titlefoot{Dipartimento di Fisica, Universit\`a di Bologna and INFN,
     Via Irnerio 46, IT-40126 Bologna, Italy
    \label{BOLOGNA}}
\titlefoot{Centro Brasileiro de Pesquisas F\'{\i}sicas, rua Xavier Sigaud 150,
     BR-22290 Rio de Janeiro, Brazil
    \label{BRASIL-CBPF}}
\titlefoot{Inst. de F\'{\i}sica, Univ. Estadual do Rio de Janeiro,
     rua S\~{a}o Francisco Xavier 524, Rio de Janeiro, Brazil
    \label{BRASIL-IFUERJ}}
\titlefoot{Coll\`ege de France, Lab. de Physique Corpusculaire, IN2P3-CNRS,
     FR-75231 Paris Cedex 05, France
    \label{CDF}}
\titlefoot{CERN, CH-1211 Geneva 23, Switzerland
    \label{CERN}}
\titlefoot{Institut de Recherches Subatomiques, IN2P3 - CNRS/ULP - BP20,
     FR-67037 Strasbourg Cedex, France
    \label{CRN}}
\titlefoot{Now at DESY-Zeuthen, Platanenallee 6, D-15735 Zeuthen, Germany
    \label{DESY}}
\titlefoot{Institute of Nuclear Physics, N.C.S.R. Demokritos,
     P.O. Box 60228, GR-15310 Athens, Greece
    \label{DEMOKRITOS}}
\titlefoot{FZU, Inst. of Phys. of the C.A.S. High Energy Physics Division,
     Na Slovance 2, CZ-182 21, Praha 8, Czech Republic
    \label{FZU}}
\titlefoot{Dipartimento di Fisica, Universit\`a di Genova and INFN,
     Via Dodecaneso 33, IT-16146 Genova, Italy
    \label{GENOVA}}
\titlefoot{Institut des Sciences Nucl\'eaires, IN2P3-CNRS, Universit\'e
     de Grenoble 1, FR-38026 Grenoble Cedex, France
    \label{GRENOBLE}}
\titlefoot{Helsinki Institute of Physics and Department of Physical Sciences,
     P.O. Box 64, FIN-00014 University of Helsinki, 
     \indent~~Finland
    \label{HELSINKI}}
\titlefoot{Joint Institute for Nuclear Research, Dubna, Head Post
     Office, P.O. Box 79, RU-101 000 Moscow, Russian Federation
    \label{JINR}}
\titlefoot{Institut f\"ur Experimentelle Kernphysik,
     Universit\"at Karlsruhe, Postfach 6980, DE-76128 Karlsruhe,
     Germany
    \label{KARLSRUHE}}
\titlefoot{Institute of Nuclear Physics PAN,Ul. Radzikowskiego 152,
     PL-31142 Krakow, Poland
    \label{KRAKOW1}}
\titlefoot{Faculty of Physics and Nuclear Techniques, University of Mining
     and Metallurgy, PL-30055 Krakow, Poland
    \label{KRAKOW2}}
\titlefoot{Universit\'e de Paris-Sud, Lab. de l'Acc\'el\'erateur
     Lin\'eaire, IN2P3-CNRS, B\^{a}t. 200, FR-91405 Orsay Cedex, France
    \label{LAL}}
\titlefoot{School of Physics and Chemistry, University of Lancaster,
     Lancaster LA1 4YB, UK
    \label{LANCASTER}}
\titlefoot{LIP, IST, FCUL - Av. Elias Garcia, 14-$1^{o}$,
     PT-1000 Lisboa Codex, Portugal
    \label{LIP}}
\titlefoot{Department of Physics, University of Liverpool, P.O.
     Box 147, Liverpool L69 3BX, UK
    \label{LIVERPOOL}}
\titlefoot{Dept. of Physics and Astronomy, Kelvin Building,
     University of Glasgow, Glasgow G12 8QQ, UK
    \label{GLASGOW}}
\titlefoot{LPNHE, IN2P3-CNRS, Univ.~Paris VI et VII, Tour 33 (RdC),
     4 place Jussieu, FR-75252 Paris Cedex 05, France
    \label{LPNHE}}
\titlefoot{Department of Physics, University of Lund,
     S\"olvegatan 14, SE-223 63 Lund, Sweden
    \label{LUND}}
\titlefoot{Universit\'e Claude Bernard de Lyon, IPNL, IN2P3-CNRS,
     FR-69622 Villeurbanne Cedex, France
    \label{LYON}}
\titlefoot{Dipartimento di Fisica, Universit\`a di Milano and INFN-MILANO,
     Via Celoria 16, IT-20133 Milan, Italy
    \label{MILANO}}
\titlefoot{Dipartimento di Fisica, Univ. di Milano-Bicocca and
     INFN-MILANO, Piazza della Scienza 3, IT-20126 Milan, Italy
    \label{MILANO2}}
\titlefoot{IPNP of MFF, Charles Univ., Areal MFF,
     V Holesovickach 2, CZ-180 00, Praha 8, Czech Republic
    \label{NC}}
\titlefoot{NIKHEF, Postbus 41882, NL-1009 DB
     Amsterdam, The Netherlands
    \label{NIKHEF}}
\titlefoot{National Technical University, Physics Department,
     Zografou Campus, GR-15773 Athens, Greece
    \label{NTU-ATHENS}}
\titlefoot{Physics Department, University of Oslo, Blindern,
     NO-0316 Oslo, Norway
    \label{OSLO}}
\titlefoot{Dpto. Fisica, Univ. Oviedo, Avda. Calvo Sotelo
     s/n, ES-33007 Oviedo, Spain
    \label{OVIEDO}}
\titlefoot{Department of Physics, University of Oxford,
     Keble Road, Oxford OX1 3RH, UK
    \label{OXFORD}}
\titlefoot{Dipartimento di Fisica, Universit\`a di Padova and
     INFN, Via Marzolo 8, IT-35131 Padua, Italy
    \label{PADOVA}}
\titlefoot{Rutherford Appleton Laboratory, Chilton, Didcot
     OX11 OQX, UK
    \label{RAL}}
\titlefoot{Dipartimento di Fisica, Universit\`a di Roma II and
     INFN, Tor Vergata, IT-00173 Rome, Italy
    \label{ROMA2}}
\titlefoot{Dipartimento di Fisica, Universit\`a di Roma III and
     INFN, Via della Vasca Navale 84, IT-00146 Rome, Italy
    \label{ROMA3}}
\titlefoot{DAPNIA/Service de Physique des Particules,
     CEA-Saclay, FR-91191 Gif-sur-Yvette Cedex, France
    \label{SACLAY}}
\titlefoot{Instituto de Fisica de Cantabria (CSIC-UC), Avda.
     los Castros s/n, ES-39006 Santander, Spain
    \label{SANTANDER}}
\titlefoot{Inst. for High Energy Physics, Serpukov
     P.O. Box 35, Protvino, (Moscow Region), Russian Federation
    \label{SERPUKHOV}}
\titlefoot{J. Stefan Institute, Jamova 39, SI-1000 Ljubljana, Slovenia
    \label{SLOVENIJA1}}
\titlefoot{Laboratory for Astroparticle Physics,
     University of Nova Gorica, Kostanjeviska 16a, SI-5000 Nova Gorica, Slovenia
    \label{SLOVENIJA2}}
\titlefoot{Department of Physics, University of Ljubljana,
     SI-1000 Ljubljana, Slovenia
    \label{SLOVENIJA3}}
\titlefoot{Fysikum, Stockholm University,
     Box 6730, SE-113 85 Stockholm, Sweden
    \label{STOCKHOLM}}
\titlefoot{Dipartimento di Fisica Sperimentale, Universit\`a di
     Torino and INFN, Via P. Giuria 1, IT-10125 Turin, Italy
    \label{TORINO}}
\titlefoot{INFN,Sezione di Torino and Dipartimento di Fisica Teorica,
     Universit\`a di Torino, Via Giuria 1,
     IT-10125 Turin, Italy
    \label{TORINOTH}}
\titlefoot{Dipartimento di Fisica, Universit\`a di Trieste and
     INFN, Via A. Valerio 2, IT-34127 Trieste, Italy
    \label{TRIESTE}}
\titlefoot{Istituto di Fisica, Universit\`a di Udine and INFN,
     IT-33100 Udine, Italy
    \label{UDINE}}
\titlefoot{Univ. Federal do Rio de Janeiro, C.P. 68528
     Cidade Univ., Ilha do Fund\~ao
     BR-21945-970 Rio de Janeiro, Brazil
    \label{UFRJ}}
\titlefoot{Department of Radiation Sciences, University of
     Uppsala, P.O. Box 535, SE-751 21 Uppsala, Sweden
    \label{UPPSALA}}
\titlefoot{IFIC, Valencia-CSIC, and D.F.A.M.N., U. de Valencia,
     Avda. Dr. Moliner 50, ES-46100 Burjassot (Valencia), Spain
    \label{VALENCIA}}
\titlefoot{Institut f\"ur Hochenergiephysik, \"Osterr. Akad.
     d. Wissensch., Nikolsdorfergasse 18, AT-1050 Vienna, Austria
    \label{VIENNA}}
\titlefoot{Inst. Nuclear Studies and University of Warsaw, Ul.
     Hoza 69, PL-00681 Warsaw, Poland
    \label{WARSZAWA}}
\titlefoot{Now at University of Warwick, Coventry CV4 7AL, UK
    \label{WARWICK}}
\titlefoot{Fachbereich Physik, University of Wuppertal, Postfach
     100 127, DE-42097 Wuppertal, Germany \\
\noindent
{$^\dagger$~deceased}
    \label{WUPPERTAL}}
\addtolength{\textheight}{-10mm}
\addtolength{\footskip}{5mm}
\clearpage

\headsep 30.0pt
\end{titlepage}

%
\pagenumbering{arabic}                  
\setcounter{footnote}{0}                %
\large
\section{Introduction}\label{s:intro}

Pentaquark is a name given to describe a bound state of four quarks and
one antiquark, e.g. $uudd\overline{s}$. The quark model does not exclude such
states. Several models predict the multiplet structure and characteristics of
pentaquarks, for example the chiral soliton model, the uncorrelated and correlated
quark models, the thermal model, lattice QCD etc.~\cite{pqpred}. The current
theoretical description of possible pentaquarks is very rich, but it does not provide a
unique picture of the pentaquark characteristics. Furthermore,
lattice calculations give very different predictions as to whether pentaquarks
exist and, if they do, what mass and parity they have.

Pentaquark states were first searched for in the 60's but the few, low
statistics, published candidates were never confirmed~\cite{confirm}.
More recent experimental evidence \cite{pqsper}, however,
may suggest the existence of pentaquark systems. The first possible candidate
is\footnote{Charge conjugated states are implied
throughout this paper.} the
$\Theta (1540)^+$, with mass of ($1.54 \pm 0.01$) GeV/$c^2$, width smaller
than 1 MeV/$c^2$, and strangeness S=+1,
consistent with being made of the quarks $uudd\bar{s}$. This evidence is still controversial
as is that for the other pentaquark states discussed in this
Letter (see \cite{revpen} and references therein).

Subsequently, evidence for another exotic baryon, doubly charged and with
double strangeness, the $\Xi(1862)^{--}$ 
(subsequently called $\Phi(1860)^{--}$, see \cite{pdg}), has been claimed by the CERN
experiment NA49~\cite{na49}, with mass of (1862 $\pm$ 2) MeV/$c^2$.

Later, the DESY experiment H1 has reported a signal for a charmed
exotic baryon in the pD$^{*-}$ channel~\cite{h1}, the ${\Theta_c(3100)^{0}}$.
This resonance was reported to have a mass of ($3099 \pm 3$ (stat) $\pm 5$
(syst)) MeV/$c^2$ and a measured width compatible with the experimental
resolution. It was interpreted as an anti-charmed baryon with a minimal
constituent quark composition of $uudd\bar c$.
Several experiments tried to verify this finding \cite{revpen}.
The ZEUS collaboration for instance challenged the results of H1;
even with a larger sample of D$^{*\pm}$ mesons, such a narrow resonance was
not observed \cite{zeus}.

Isospins 0 and 1 are both possible for pentaquarks; isospin 1 would lead to
three charge states $\Theta^0$, $\Theta^+$ and $\Theta^{++}$. Thus the search
is for a family of pentaquarks.

This paper reports on the results of a search for
pentaquark states in hadronic Z decays recorded by DELPHI.
In a similar analysis, ALEPH \cite{aleph} did not
observe significant signals. The powerful particle identification characterizing the DELPHI
detector might facilitate this search, since this feature helps in detecting and separating from the background some
decay states of pentaquarks.

The article is organised as follows. After a short description of the
subdetectors used for the analysis (Section 2), Section 3 presents the results
of a search for pentaquarks in the pK$^0$ (the $\Theta^+$) and the pK$^+$ (the $\Theta^{++}$)
channels. Section 4 presents a search for a doubly-charged, doubly-strange
pentaquark (the $\Phi(1860)^{--}$). Section 5 presents a search for a charmed
pentaquark (the ${\Theta_c(3100)^{0}}$). A summary is given in Section 6.

\section{The Detector}

The DELPHI detector is described in detail in~\cite{detector}, and its
performance is analysed in \cite{perfo}. 

The present analysis relies mostly on information provided by
the central tracking detectors and the {Barrel Ring Imaging Cherenkov Counter}
(BRICH):
\begin{itemize}
\item The {microVertex Detector} (VD) consists of three layers of silicon strip
      detectors at radii\footnote{In the standard DELPHI coordinate system, 
the $z$ axis is along the electron beam direction, the $x$ axis 
points towards the center of LEP, and the $y$ axis points upwards. 
The polar angle to the $z$ axis is called $\theta$ and the azimuthal 
angle around the $z$ axis is called $\phi$; 
the radial coordinate is $R = \sqrt{x^2+y^2}$.} of 6.3~cm, 9.0~cm and 10.9~cm. 
       $R\phi$ is measured in all three layers. The first and third layers also
      provide $z$ information
      (from 1994 on).
      The $\theta$ coverage for a
      particle passing all three layers is from $44^\circ$ to $136^\circ$.
      The single point precision has been estimated from real data to be
      about 8 $\mu$m in $R\phi$ and (for charged particles crossing
      perpendicular to the module) about 9 $\mu$m in $z$.
\item The {Inner Detector} (ID) consists of an inner drift chamber with
      jet chamber geometry and 5 cylindrical MWPC (straw tube from 1995 on)
      layers. The jet chamber,
      between 12 and 23~cm in $R$ and from $23^\circ$ to $157^\circ$ in
      $\theta$
      ($15^\circ$-$165^\circ$ from 1995 on),  consists of 24 azimuthal sectors, each providing up to
      24 $R\phi$ points.
\item The {Time Projection Chamber} (TPC) is the main tracking device. It
      provides up to 16 space points per particle trajectory
      for radii between 40~cm and 110~cm. The precision on the track
      elements is about 150 $\mu$m in $R\phi$ and about 600 $\mu$m in $z$.
      A measurement of the specific energy loss $dE/dx$ of a track is provided with
      a resolution of about 6.5\%, providing charged particle identification
      up to a momentum of about 1 GeV/$c$.
\item The {Outer Detector} (OD) is a 4.7~m long set of 5 layers of drift tubes
      situated at 2~m radius to the beam which provides precise spatial
      information in $R\phi$.
\item The {Barrel Ring Imaging Cherenkov Counter} (BRICH) is
      the main DELPHI detector devoted to charged particle identification.
      It is subdivided into two halves ($ z > 0$ and $z < 0$) and provides
      particle identification using Cherenkov radiation produced in a
      liquid or a gas radiator. This radiation, after appropriate focusing, is
      transformed into photoelectrons in a TPC-like drift structure and the
      Cherenkov angles of the track in both media are determined.
      The BRICH provides particle identification in the momentum range from
      $0.7$ GeV/$c$ to $45$ GeV/$c$.
\end{itemize}
The DELPHI tracking system was completed by two 
tracking chambers (FCA and FCB) in each forward region.

To compute the selection efficiency of the various channels studied,
Z~$\rightarrow q\overline{q}$ events were simulated using the JETSET
parton shower generator \cite{jetset} and then processed through the DELPHI
simulation program, DELSIM, which models the detector response.
The simulated events passed through DELSIM were then processed by the same
reconstruction program as used for the data, DELANA~\cite{perfo}. The amount
of simulated events is more than twice the real data.

For the $\Theta^+$, $\Theta^{++}$ and $\Phi(1860)^{--}$ searches, the data recorded during the LEP1 operation in the years  1991 to 1995 were used. For the $\Theta_c(3100)^0$ search, the analysis was restricted to the years 1994 and 1995, the two
highest luminosity years of LEP1, with all DELPHI particle identifiers fully operational.

\section{Search for Strange Pentaquarks in the pK system}
The state $\Theta^+$ can be detected through its decay into pK$^0$ pairs; the
state $\Theta^{++}$ could be detected through its decay into pK$^+$.
Therefore the invariant mass distributions of pK$^0$ and pK$^+$
pairs in hadronic Z decays were studied. These were compared with the pK$^-$ spectrum,
where the $\Lambda$(1520) is observed.

\subsection{Event selection\label{s:det_ana}}

Hadronic Z decays for this analysis were selected by requiring at least four
reconstructed charged particles and a total energy of these particles (assuming
the pion mass) larger than $12\%$ of the centre-of-mass (c.m.) energy.
The charged-particle tracks had to be longer than 30 cm, with
a momentum larger than 400 MeV/$c$ and a polar angle between $20^\circ$ and
$160^\circ$. The polar angle of the thrust axis, $\theta_{thrust}$, was
computed for each event and events were rejected if $|\cos\theta_{thrust}|$
was greater than 0.95. A total of 3.4 million hadronic events were selected.

The  selection efficiency for hadronic events was estimated  using the simulation, and
found to be larger than $95\%$ within the angular acceptance.

In order to search for the pentaquark states, the pK$^0$,
pK$^-$ and pK$^+$ invariant mass spectra were constructed
using identified particles. Particle identification was performed combining
$dE/dx$ and BRICH information. According to the quality of particle
identification the tagging categories loose, standard and tight  are
distinguished for each particle species as well as for so-called
``heavy'' tag, which  severely reduces the fraction of charged pions.
To further improve the quality of particle identification for a track of given
momentum and (assumed) particle type it was required that information
from the detectors specified in Table~\ref{t:id-ranges} was present. Only in the years 1994 and 1995 all 
particle identification detectors were fully operational; the identification was essentially only based on TPC during the years  1991 to 1993.

\newcommand{\mc}{\multicolumn}
\begin{table}[tbh]
{
\begin{center}{
\begin{tabular}{|c|c|c|c|c|c|c|c|}
   \hline
   & \multicolumn{7}{c|}{momentum range in~GeV/$c$}\\
  \cline{2-8}
   & 0.3 - 0.7 & 0.7 - 0.9 & 0.9 - 1.3 & 1.3 - 2.7 & 2.7 - 9.0 &
                                         9.0 -16.0 & 16.0 - 45.0 \\
  \hline
  \hline
       &     & \mc{3}{c|}{}        & \mc{3}{c|}{}\\
 $\pi$ & TPC & \mc{3}{c|}{LRICH S} & \mc{3}{c|}{GRICH S}\\
       &     & \mc{3}{c|}{}        & \mc{3}{c|}{}\\
  \hline
       &     & \mc{3}{c|}{}        & GRICH V &\mc{2}{c|}{}\\
   K   & TPC & \mc{3}{c|}{LRICH S} &    +    &\mc{2}{c|}{GRICH S}\\
       &     & \mc{3}{c|}{}        & LRICH S &\mc{2}{c|}{}\\
  \hline
       & \mc{2}{c|}{}    &  TPC    &         & GRICH V &         &         \\
   p   & \mc{2}{c|}{TPC} &  +      & LRICH S &    +    & GRICH V & GRICH S \\
       & \mc{2}{c|}{}    & LRICH V &         & LRICH S &         &         \\
  \hline
\end{tabular}}
\end{center}
\caption[]{Momentum ranges for particle identification: TPC denotes
         identification using the $dE/dx$ measurement of the TPC,
         LRICH S (V) denotes identification using a signal (veto) of
         the liquid RICH, and correspondingly GRICH for the gas RICH.
}
\label{t:id-ranges}
}
\end{table}

A particle was taken to be a proton if it was tightly tagged or fulfilled the standard tag by
identification from ionization loss in the TPC.
Kaons were required to be tightly tagged in the momentum ranges
$p<3.5$~GeV/$c$ and $p>9.5$~GeV/$c$. In the intermediate momentum range kaons
were also identified by a tight heavy particle tag \cite{emile} combined with
at least a standard kaon tag.

\subsubsection{Description of the invariant mass spectra}
In the present analysis, the mass spectra were described by a
distribution function,  $f(M,\vec{a})$, of the invariant mass $M$.
The parameters $\vec{a}$
were determined by a least squares fit of the function to the data.
The function $f(M,\vec{a})$ was composed of two parts:
\begin{equation}
   f(M,\vec{a}) = f^{S}(M,\vec{a}) + f^{B}(M,\vec{a})
                                     \, ,
                                     \end{equation}
corresponding to the signal and to the background respectively.
The signal function, $f^{S}(M,\vec{a})$, described the resonance signals
in the corresponding invariant mass distributions. It has the form:
\begin{eqnarray}
f^S(M,\vec{a}) = a_1 \times R(M,a_2, a_3),
\end{eqnarray}
where $R$ is either a non-relativistic Breit-Wigner or a normalised Gaussian
function accounting for the resonance production; $a_2$ and $a_3$ are
respectively the fitted peak RMS width and mass $m$.
The background term, $f^{B}(M,\vec{a})$, was taken to be a third order
polynomial in $M$.


\subsection{Analysis of the pK$^0$ channel\label{s:K0}}
The invariant mass distribution for pK$^0$ pairs was first studied. K$^0$
candidates were obtained from the fit of charged particle tracks of opposite charge
 consistent with the pion hypothesis, as described in \cite{perfo}.
The $\pi^+\pi^-$ invariant mass is shown in Figure 1 (a).

The pK$^0_S$ mass distribution is displayed in Figure 1 (b), for an invariant K$^0_S$ mass between 0.45 GeV/$c^2$ and 0.55 GeV/$c^2$. No signal is
visible in the $\Theta^{+}$ mass region; the simulation accounts
very well for the data over the whole mass spectrum.

\begin{figure}
\center{\epsfig{file=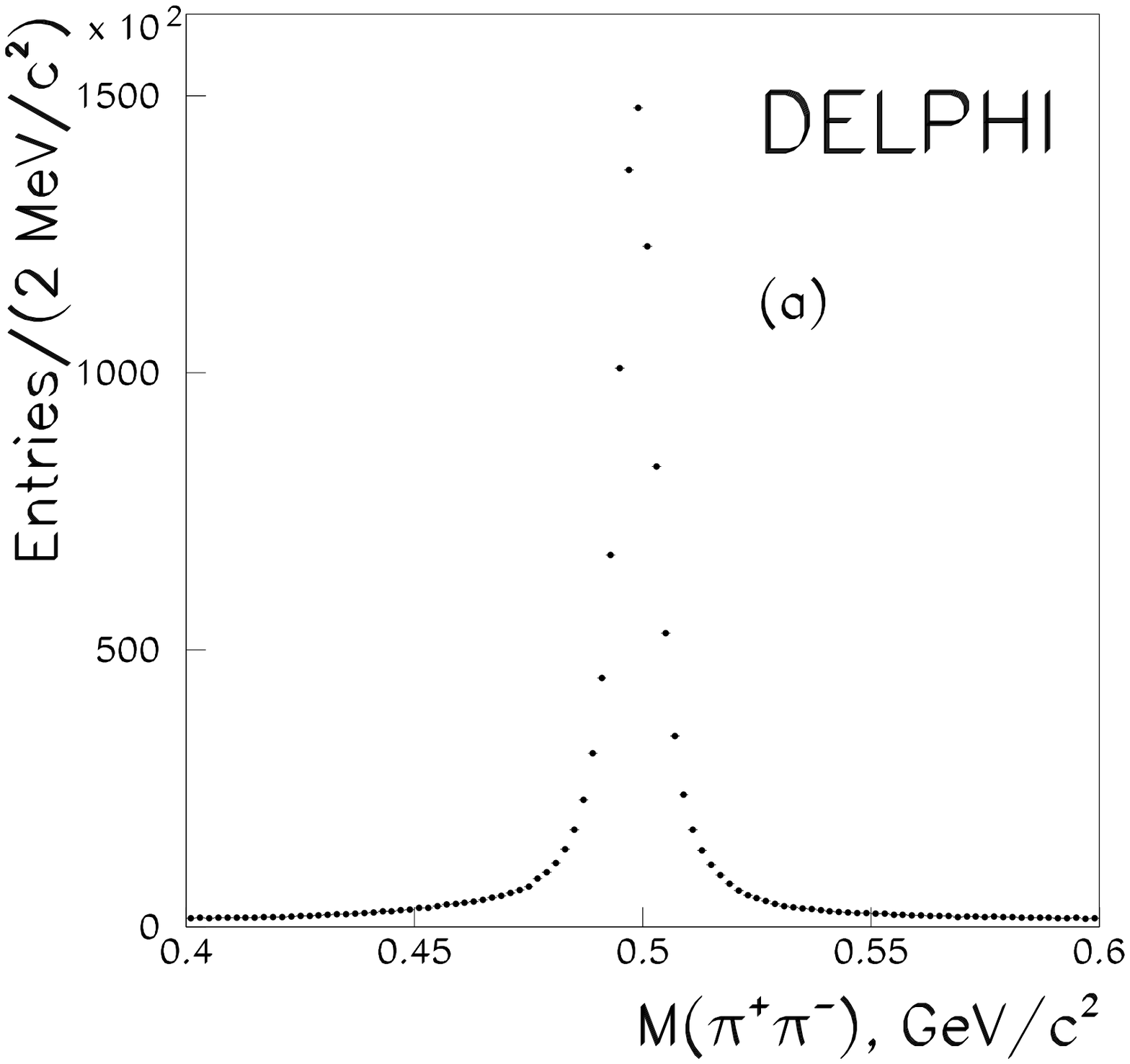,width=0.48\textwidth}\epsfig{file=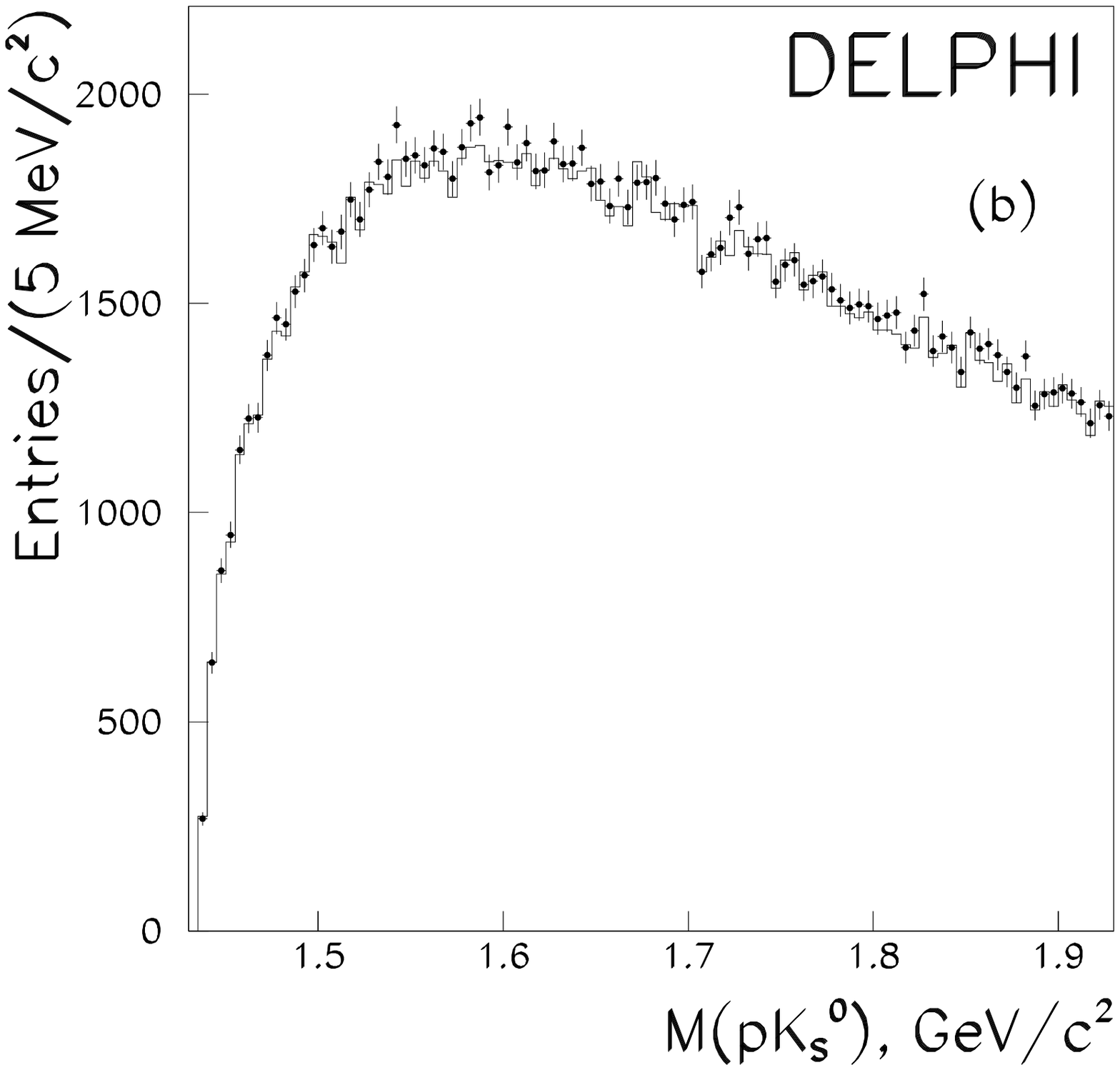,width=0.48\textwidth}}
\caption[thetaplus]
{
\label{f:thetapmass}
(a) $\pi^+\pi^-$ invariant mass (b) pK$^0$ mass spectrum. The histogram represents the simulation, while the points represent the data.
}
\end{figure}

To set the limit on the $\Theta^{+}$ production, the fitting
procedure as described above was applied, modeling a possible signal by a Gaussian
function with a central value of \mbox{1.54 GeV/$c^2$} and a RMS 
width\footnote{Throughout the paper, if the fit is done by a Breit-Wigner function
the width indicates the value of the $\Gamma$ parameter, while in the case of a Gaussian function it indicates the RMS error $\sigma$.} of 10 MeV/$c^2$, equal to the resolution.

The pK$^0_S$ selection efficiency was
estimated  from a  Monte Carlo
generated sample of $\Theta^{+}$ events to be $(6.4 \pm 0.3)\%$. The error is dominated by the systematic uncertainties coming from K$^0_S$
reconstruction and proton identification.

The estimated number of events in the signal region was $-20 \pm 64$ (stat).
The corresponding
upper limit, at 95\% CL, on the average production rate per hadronic event of the
$\Theta^{+}$ is: \[ \langle N_{\Theta^{+}} \rangle  \times Br(\Theta^+ \rightarrow pK^0_S)   \; \; <  5.1 \times 10^{-4} \, , \]
where the systematic uncertainty was added in quadrature to the statistical error. The result has been corrected for the 
branching fraction $Br(K^0_S \rightarrow \pi^+\pi^-)$.

\subsection{Analysis of the pK$^-$ and pK$^+$ channels\label{s:K-+}}

The search for a possible $\Theta^{++}$ was made in the pK$^+$ channel,
after investigation of the channel pK$^-$, where the presence of the
$\Lambda$(1520) resonance allows the pK$^-$(K$^+$) selection
efficiency to be measured  in the region of interest.
Figure 2 (a) shows the pK$^-$ invariant mass spectrum. A clear
$\Lambda$(1520) signal is observed at the expected mass. It has been checked
that there are no prominent reflections from known particle decays in the
pK$^-$ mass spectrum. In addition pK$^-$ combinations in which the K$^-$
combined with any identified K$^+$ had a mass in the $\phi\ (1020)$ region
were discarded.
The total excess in the $\Lambda$(1520) region, measured from the fit to the
mass spectrum of Figure 2 (a) is of:
\begin{equation}
\langle n_{\Lambda(1520)} \rangle = 2130 \pm 450 \, \, {\rm events},
\end{equation}
with a mass of $1.520 \pm 0.002$ GeV/$c^2$ and a width of
$0.010 \pm 0.004$ GeV/$c^2$, compatible with the experimental resolution.
The $\chi^2$ per degree of freedom is 1.4.
The $\Lambda$(1520) selection efficiency determined from the simulation
is  $(12.8 \pm 0.5)\%$. This corresponds to an average $\Lambda$(1520)
production rate per hadronic event of $0.0217 \pm 0.0046$ (stat) to be compared with
the published value \cite{pdg} of \mbox{$0.0224 \pm 0.0027$}.

The invariant mass spectrum for pK$^+$ pairs, obtained using the same cuts,
is plotted in Figure 2 (b). No significant peak is visible; the $\chi^2$ per degree of
freedom of the fit to the background function only is 2.1.

\begin{figure}
\center{
\epsfig{file=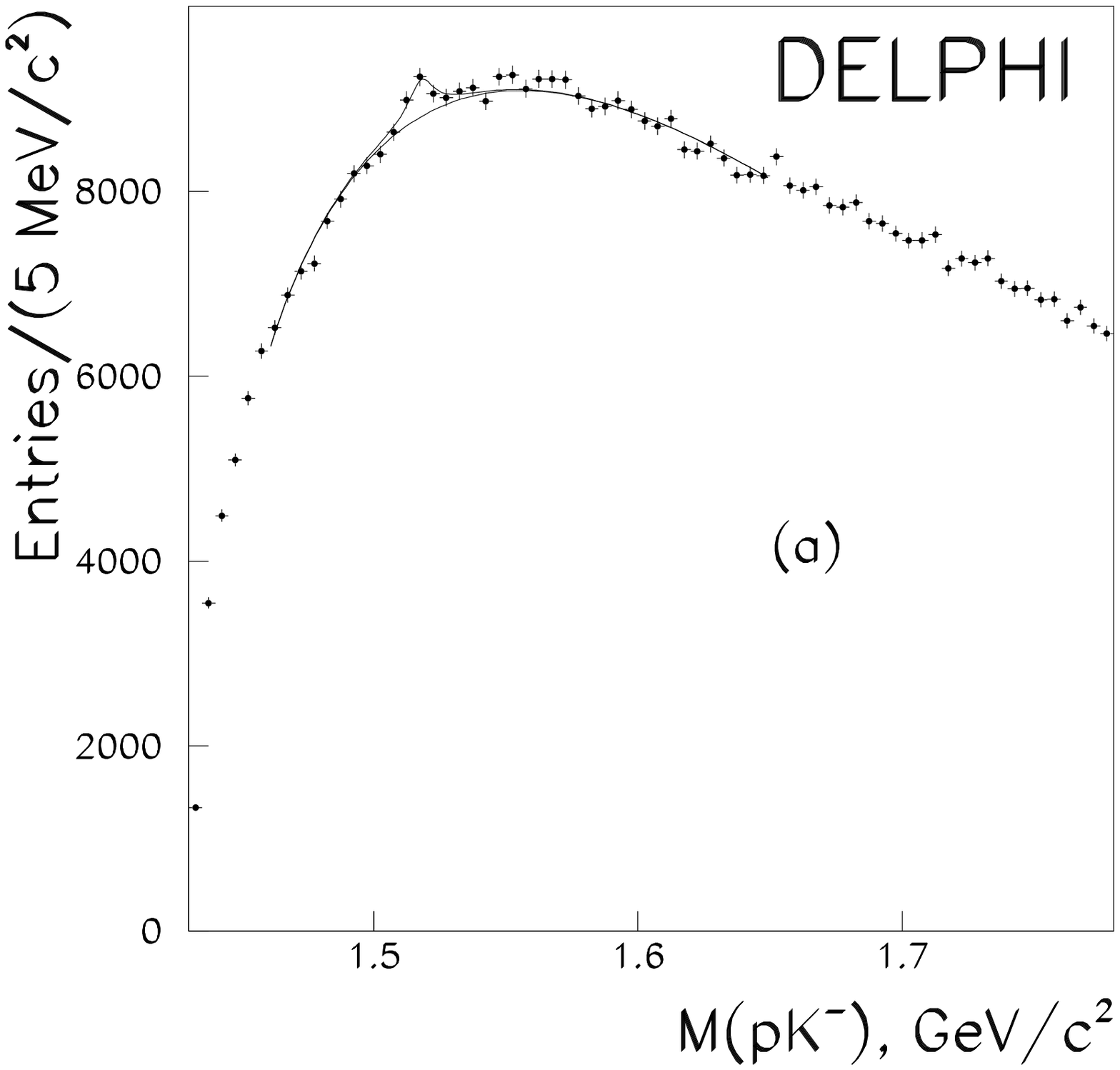,width=0.48\textwidth}\epsfig{file=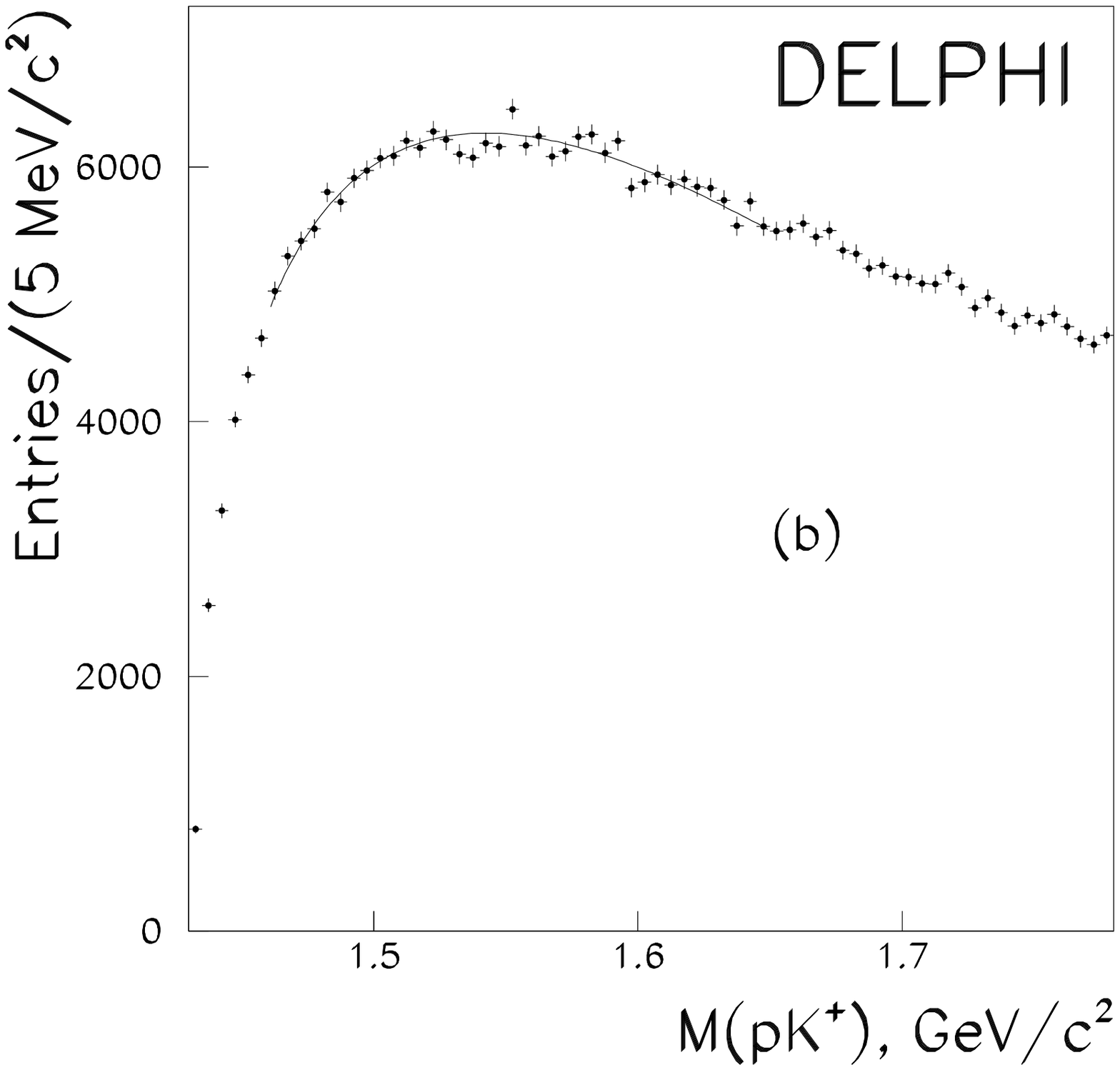,width=0.48\textwidth}}
\caption[]
{
\label{f:thetappmass}
(a) Differential pK$^-$  and (b)  pK$^+$ mass spectra.
The lines represent the fits described in the text.
}
\end{figure}

 An upper limit  for the average production rate of the $\Theta^{++}$ can be
determined over the range of its mass estimates (1.45 to 1.65 GeV/$c^2$), assuming
the same efficiency as for the $\Lambda$(1520). It should be taken into account
that, since the $\Lambda$(1520) can decay into a charged pair and into a
neutral pair as well, essentially with the same probability, the sensitivity
to decay channels of the $\Theta^{++}$ is twice that of the $\Lambda$(1520).

A fit to the form (1) was performed by varying the mass between 1.45 GeV/$c^2$
and 1.65 GeV/$c^2$ in steps of 5 MeV/$c^2$, and by imposing a RMS width of
10 MeV/$c^2$ (the expected experimental resolution). Limits at
95\% CL were then calculated as a function of the mass, yielding a maximum signal of $350 \pm 187$ events. 
The systematic uncertainties
on the production rate of a $\Theta^{++}$  with a mass close to the $\Lambda(1520)$
mass, can be expected to be of the same order as those of the $\Lambda(1520)$
production rate, which were estimated to be of 16\% \cite{l1520}. Such systematics are therefore negligible with respect to the error from the fit.
A general limit
\[ \langle N_{\Theta^{++}} \rangle \; \; <  1.6 \times 10^{-3} \, \]
for the mass region between 1.45 GeV/$c^2$ and 1.65 GeV/$c^2$ is obtained. 
This limit is higher than what could be expected given the sensitivities, 
due to the about $2\sigma$ statistical
fluctuations in the mass region between 1.52 GeV/$c^2$ and
1.58 GeV/$c^2$.




\section{Search for Doubly Charged and Doubly Strange Pentaquarks in the
$\Xi^- \pi^-$ system}


The exotic baryons
with double charge and double strangeness were searched through the decay into into $\Xi^-\pi^-$.
The hadronic Z decays sample for this analysis is the same as described in
section 3.1; it corresponds to a total of 3.4 million hadronic events  after the cuts, recorded in the years 1991 to 1995.

\subsection{$\Xi^-$ Reconstruction}

The $\Xi^-$ hyperon was reconstructed through the decay
$\Xi^- \rightarrow \Lambda \pi^-$.
For this, all $V^0$ candidates, i.e., all pairs of oppositely
charged particles, were considered as $\Lambda$ candidates.
For each pair, the higher momentum particle was assumed to be a proton and
the other a pion, and a vertex fit performed using the standard DELPHI
$V^0$ search algorithm~\cite{perfo}.

The p$\pi^-$ invariant mass is shown in Figure 3 (a).
The $\Lambda$ candidates were selected by requiring an invariant mass
$M($p$\pi^-)$ between 1.100 GeV/$c^2$ and 1.135 GeV/$c^2$, a $\chi^2$
probability of the $V^0$ vertex fit larger than $10^{-5}$ and
a decay length from the interaction point greater than 0.2 cm in the plane transverse to the beam.

A constrained multivertex fit was performed on each $\Xi^-$ candidate
decaying into  $\Lambda \pi^-$ \cite{pap175}.
The 16 measured variables in the fit were the five parameters of the helix
parameterization of each of the three charged particle tracks and
the $z$ coordinate of the beam interaction point (the $x$ and $y$
coordinates were so precisely measured that they could be taken as fixed).
The fitted variables were the decay coordinates of the $\Xi^-$ and
$\Lambda$.

The fit constrained the sum of the $\Lambda$ and $\pi$ momenta to be equal to
the $\Xi^-$ momentum. The constraint on the $\Lambda$ decay products to give
the nominal $\Lambda$ mass value $1115.683\pm0.006$ MeV/$c^2$ \cite{pdg} was
also applied.

The resulting $\Lambda \pi^-$ invariant mass spectrum after
the fit is shown in Figure 3~(b).

\begin{figure}
\center{\epsfig{file=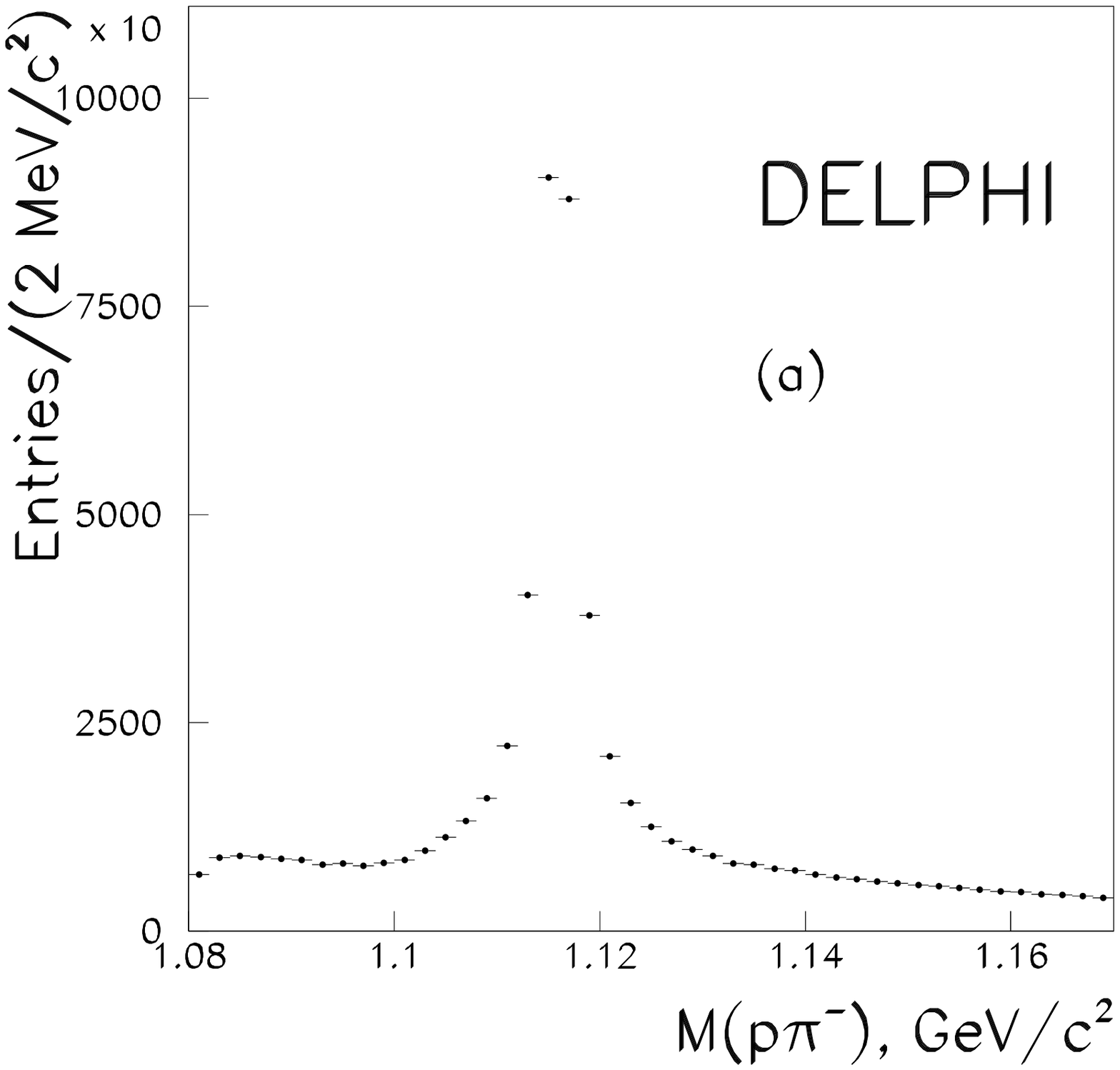,width=0.48\textwidth}\epsfig{file=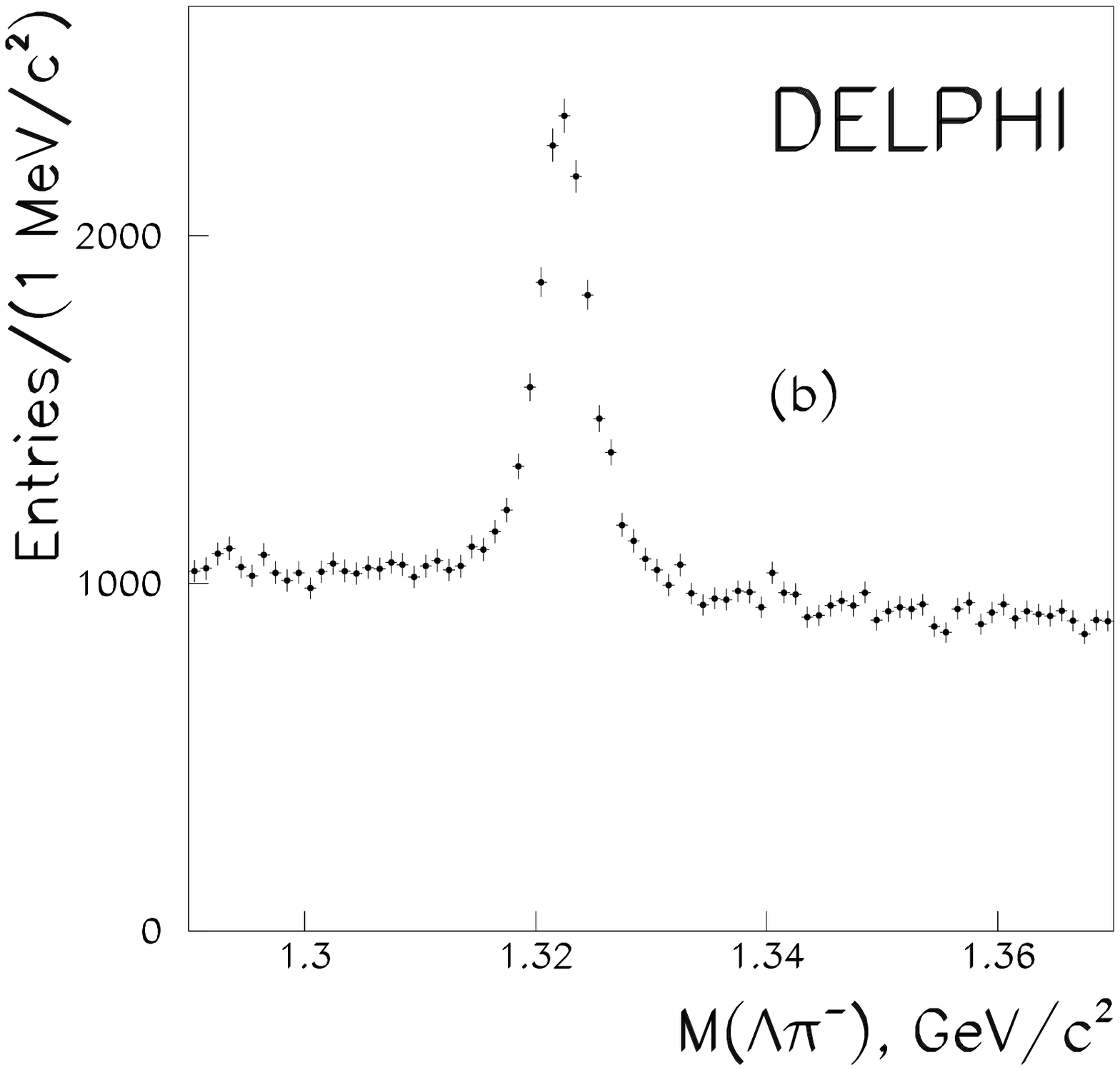,width=0.48\textwidth}}
\center{\epsfig{file=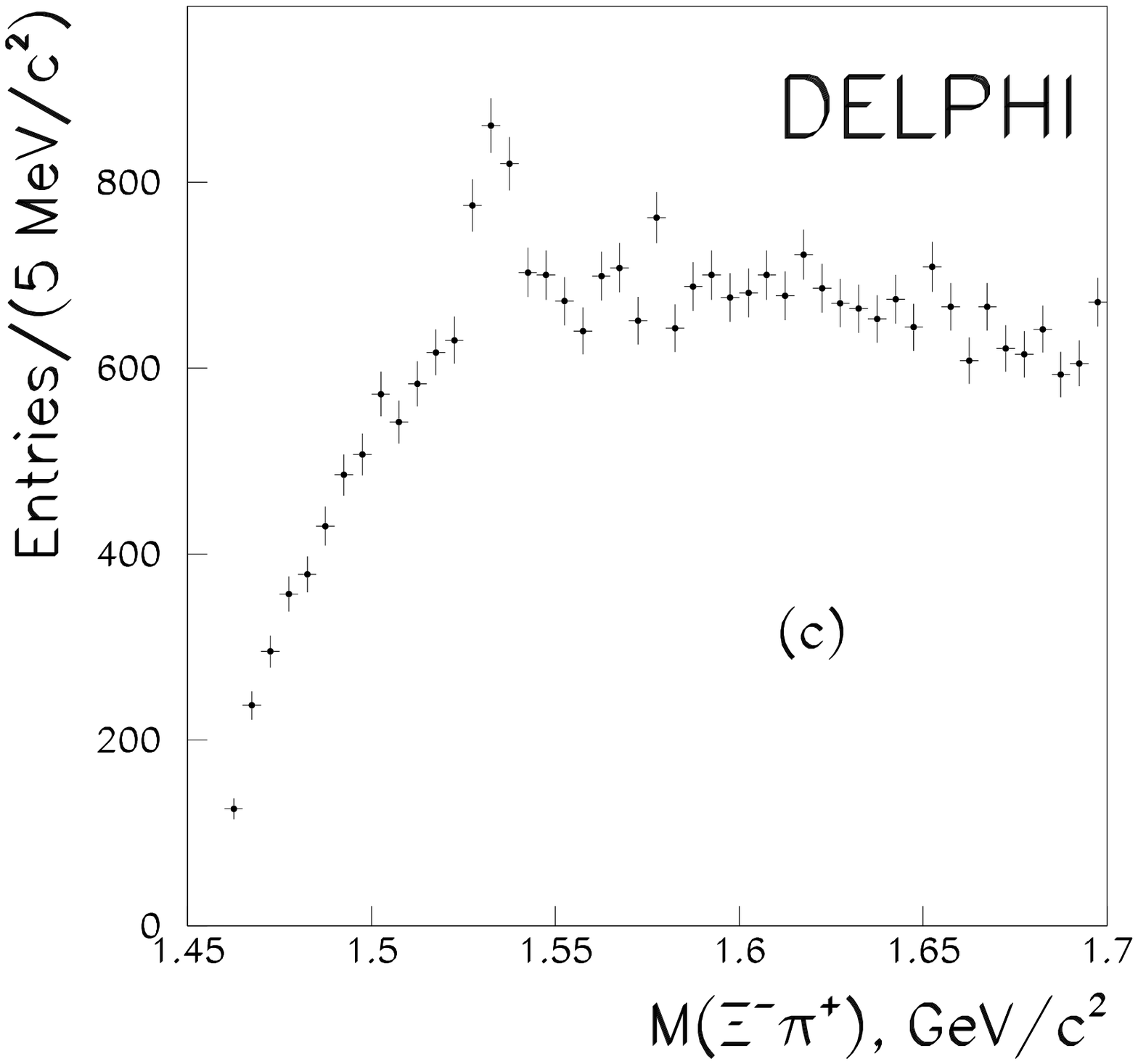,width=0.48\textwidth}\epsfig{file=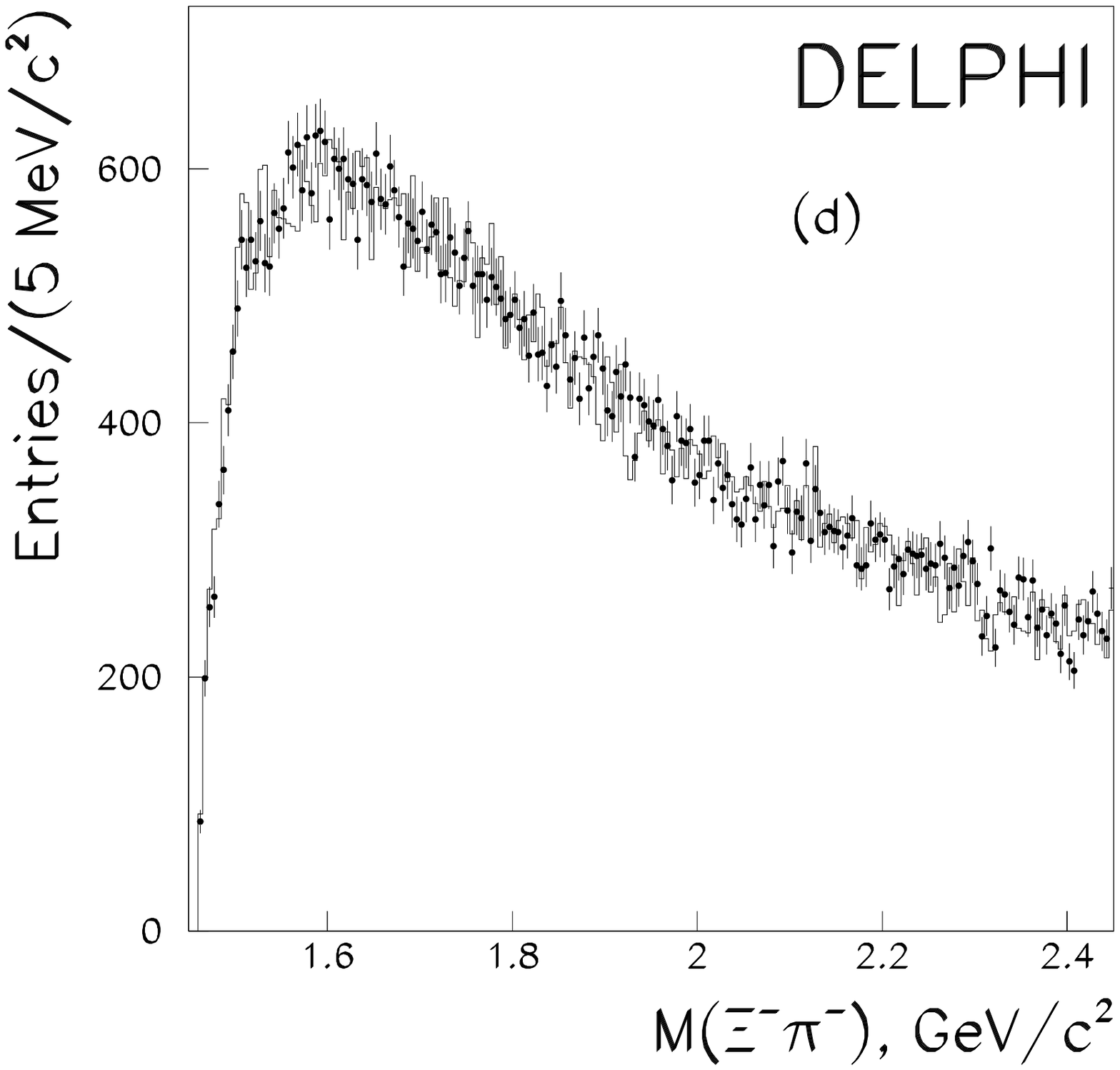,width=0.48\textwidth}}
\caption[]{(a) Invariant p$\pi^-$ mass spectrum. (b) Invariant $\Lambda \pi^-$ mass distribution.
(c) Invariant $\Xi^- \pi^+$ mass distribution. (d) Invariant $\Xi^- \pi^-$ mass distribution.
The histogram represents the simulation.}
\end{figure}

\subsection{Analysis of the $\Xi \pi$ system }

Each reconstructed $\Xi^-$ candidate in the mass range between
1.30 GeV/$c^2$ to 1.34 GeV/$c^2$ was combined with a pion.

The mass spectrum of neutral combinations $\Xi^-\pi^+$ is shown in Figure 3 (c);
a clear $\Xi(1530)$ peak of $820 \pm 50$  events is observed. The production properties of
$\Xi(1530)$ have already been measured by DELPHI in \cite{xi1530}.

The mass spectrum of combinations $\Xi^-\pi^-$ is shown in Figure 3 (d). No
significant excess is observed. The histogram shows the prediction
of the simulation for the $\Xi^-\pi^-$  spectrum without pentaquarks.
To estimate the number of pentaquarks we performed a fit of the form (1)
to the $\Xi^-\pi^-$ mass spectrum, with a Gaussian central value
of 1.862 GeV/$c^2$ and a width of 0.015 GeV/$c^2$ equal to the resolution in
this mass region. The number of events resulting from the fit is equal
to $-50\pm75$, dominated by the error from the fit itself. The reconstruction efficiency of a possible $\Phi(1860)^{--}$
object decaying into $\Xi^- \pi^-$ has been computed from a  Monte Carlo
generated sample of $\Phi(1860)^{--}$ events, to be $(10.0\pm0.5)\%$; the error is dominated by the
uncertainties on particle reconstruction and identification.
This leads to an estimate of the upper limit of the production rate of a
$\Phi(1860)^{--}$ object decaying into $\Xi^- \pi^-$ per hadronic Z  decay, 
at  $95\%$ CL:
\[ \langle N_{\Phi(1860)^{--}} \rangle \times Br(\Phi(1860)^{--} \rightarrow \Xi^- \pi^-) \; \; <  2.9 \times 10^{-4} \, . \]



\section{Search for Charmed Pentaquarks in the D$^*$p \mbox{system}}


\subsection{Event Selection}

After the standard hadronic event selection criteria listed in section 3.1
were applied to the data collected in 1994 and 1995, about 2.1 million hadronic events remained. 

Events containing the decay chain
$\mathrm{D^{*+} \rightarrow D^0 X\rightarrow K^-\pi^+ X}$
were selected as a first step of the analysis.
The following selection criteria were required to suppress the background:
\begin{itemize}
\item $x_E($K$\pi) \geq 0.15$, where $x_E$ is the energy fraction with respect
to the beam energy;
\item in the reconstructed D$^0$ decay, it was required that both the kaon and
pion momenta were larger than 1 GeV/$c$,  and that the angle between the
K and $\pi$ momenta were smaller than $90^\circ$ in the D$^*$ system;
\item the momentum of the bachelor pion (the soft pion coming from the D$^*\rightarrow$D$\pi$ decay) had to be between 0.3 GeV$/c$ and
2.5 GeV$/c$, and the angle between the bachelor $\pi$ momentum in the rest frame of the reconstructed D$^0$ and the momentum of the D$^0$ candidate had to be smaller than $90^\circ$;
\item the decay~length of the D$^0$ had to be smaller than
2.5 cm, but positive by at least three standard deviations;
\item $\cos \theta_K > -0.9$, where $\cos \theta_K$ is the angle between
the D$^0$ flight direction and the K direction in the D$^0$ rest frame;
\item  the invariant mass of the K$\pi$ system had to be between $1.79~$GeV/$c^2$ and 
1.91~GeV/$c^2$, and the mass difference $\Delta M = M_{K\pi\pi} - M_{K\pi}$
was required to be between $0.1425~$GeV/$c^2$ and 0.1485~GeV/$c^2$;
\item the K and $\pi$ candidates  were required to have at least one hit in
the VD;
\item the K candidates should not have a positive pion tag.
This requirement suppresses about 50\% of the combinatorial background
surviving all other cuts.
\end{itemize}

\begin{figure}
\center{\epsfig{file=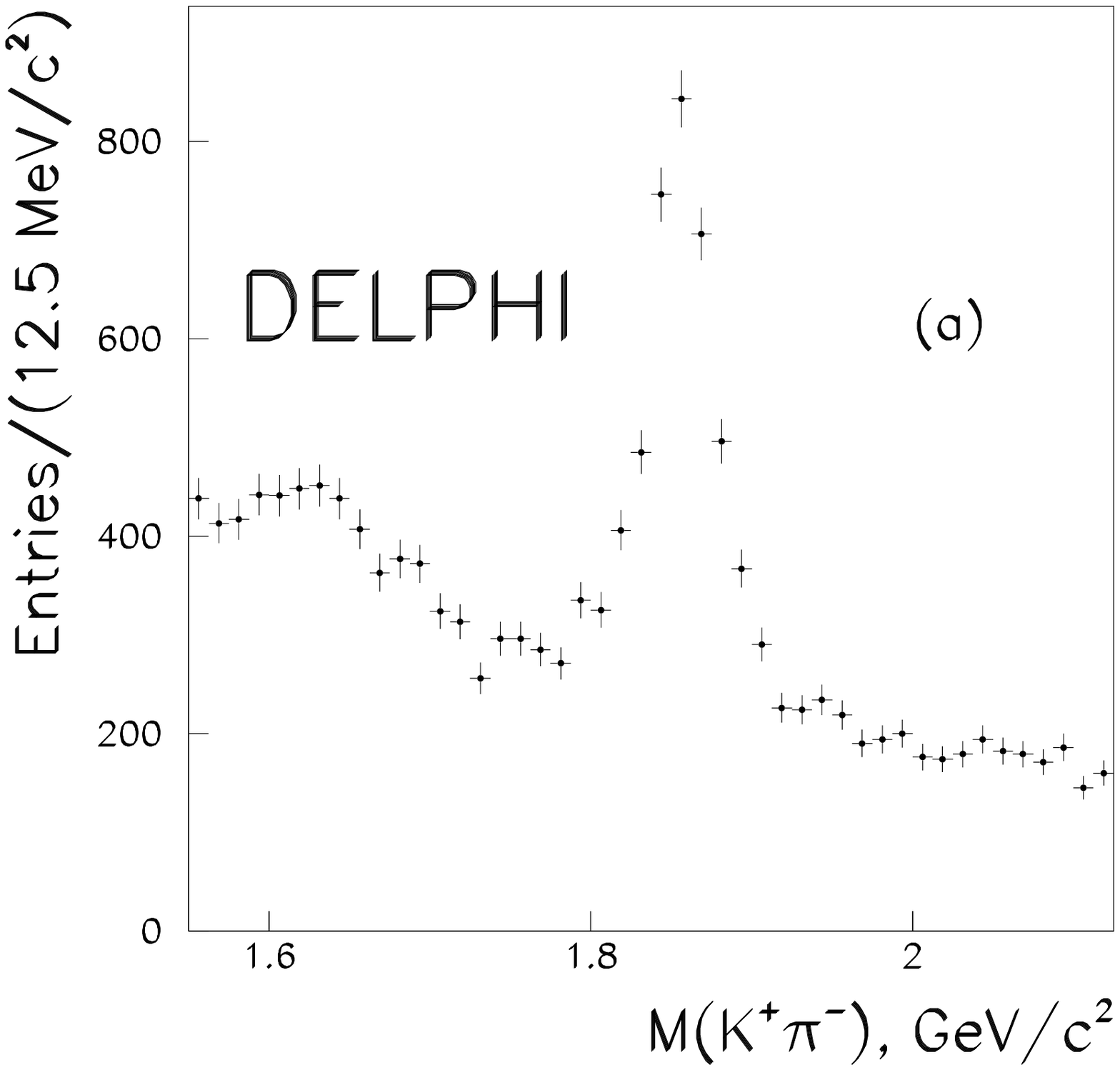,width=0.48\textwidth}\epsfig{file=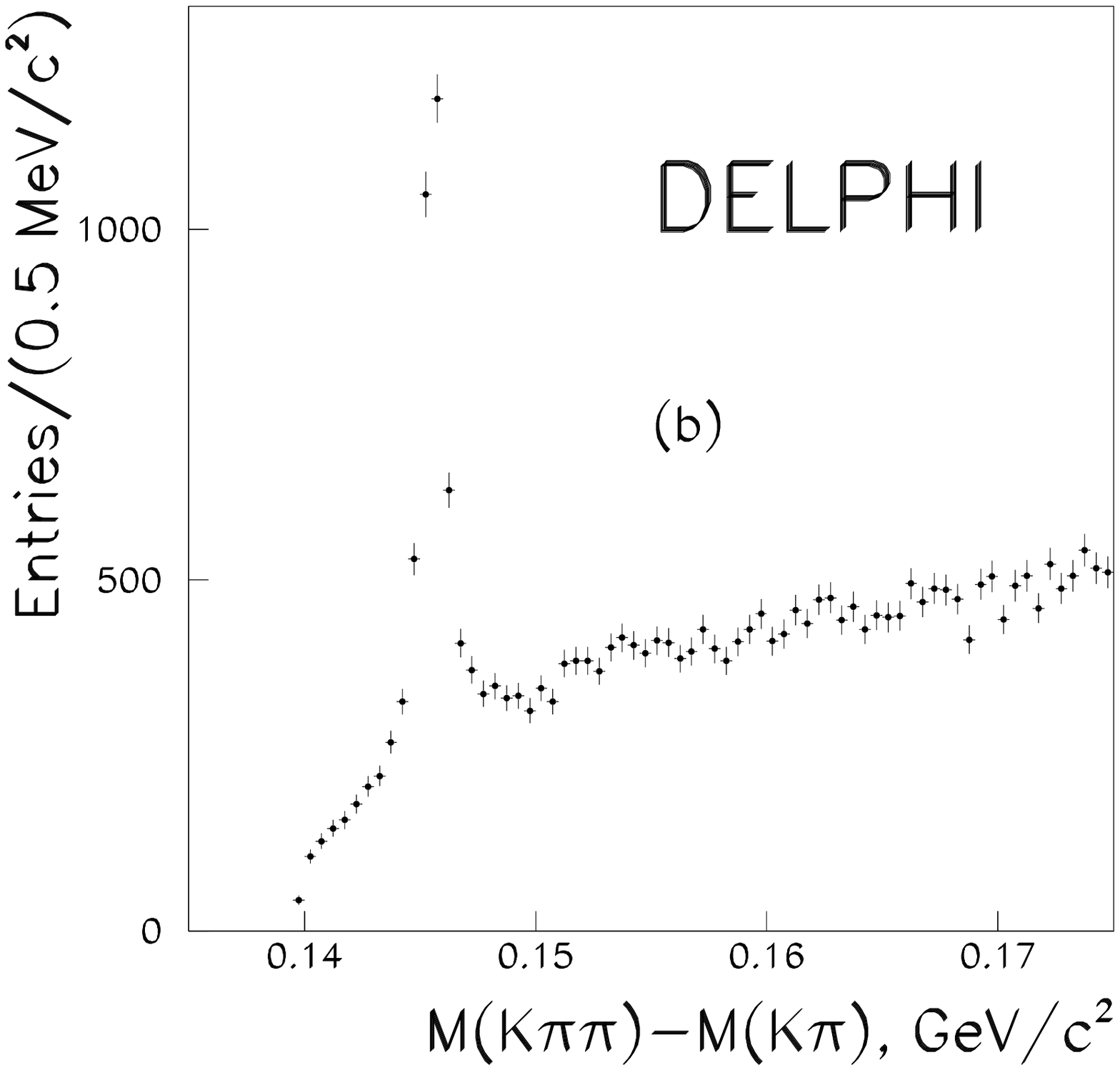,width=0.48\textwidth}}
\caption{\label{f:mkpin}(a) Invariant K$^+\pi^-$ mass. (b)
Distribution of $\Delta M = M_{K\pi\pi} - M_{K\pi}$.}
\end{figure}

\subsection{Analysis of the D$^*$p system}

The $M_{K\pi}$ and $\Delta M$ spectra obtained after the cuts
listed above are shown in Figure \ref{f:mkpin}.
The backgrounds
around the very clear D$^0$ and D$^*$ (corresponding to the decay
D$^* \rightarrow$~D$^0\pi$) peaks are quite small. 

\begin{figure}
\center{\epsfig{file=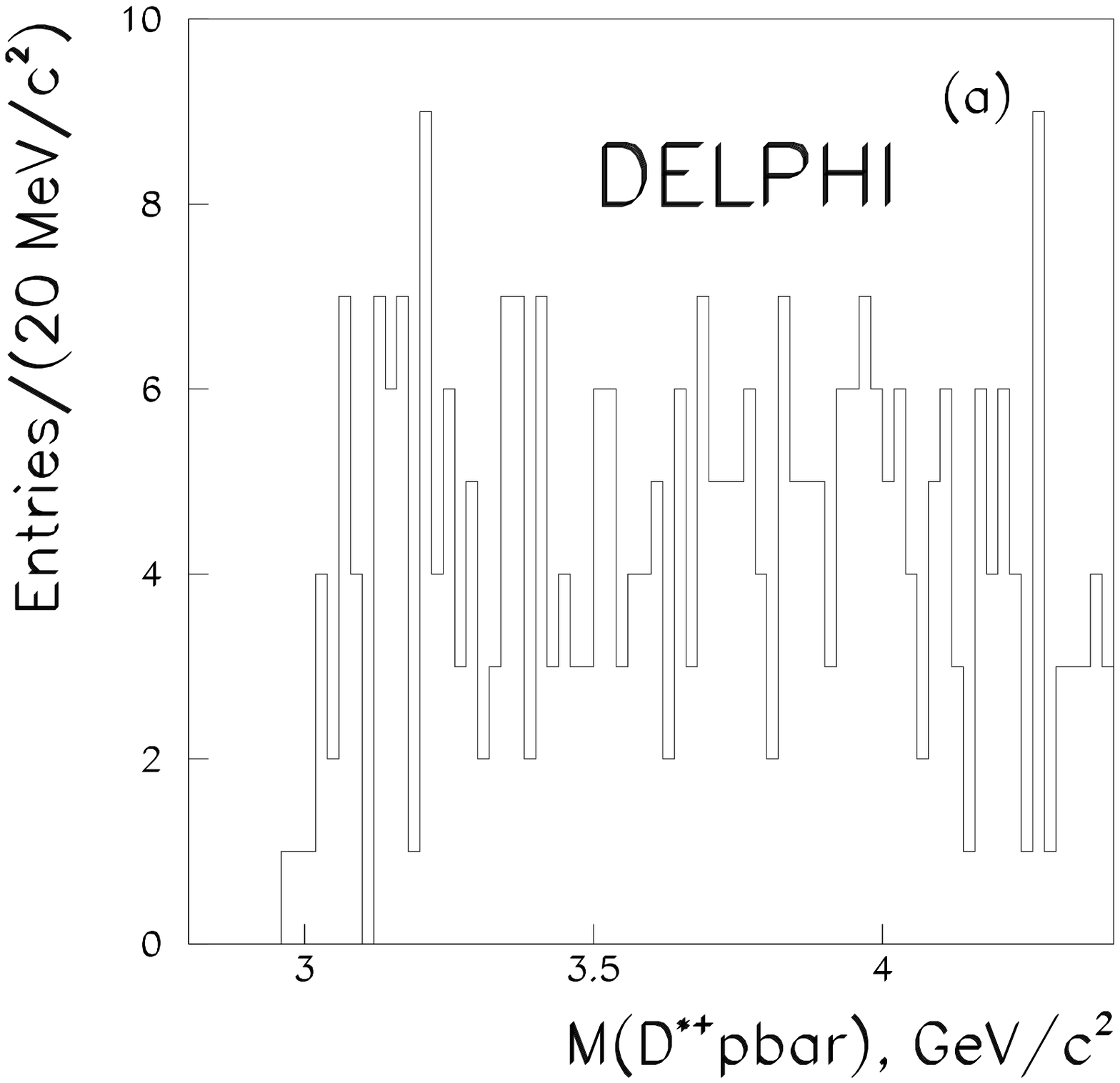,width=0.48\textwidth}\epsfig{file=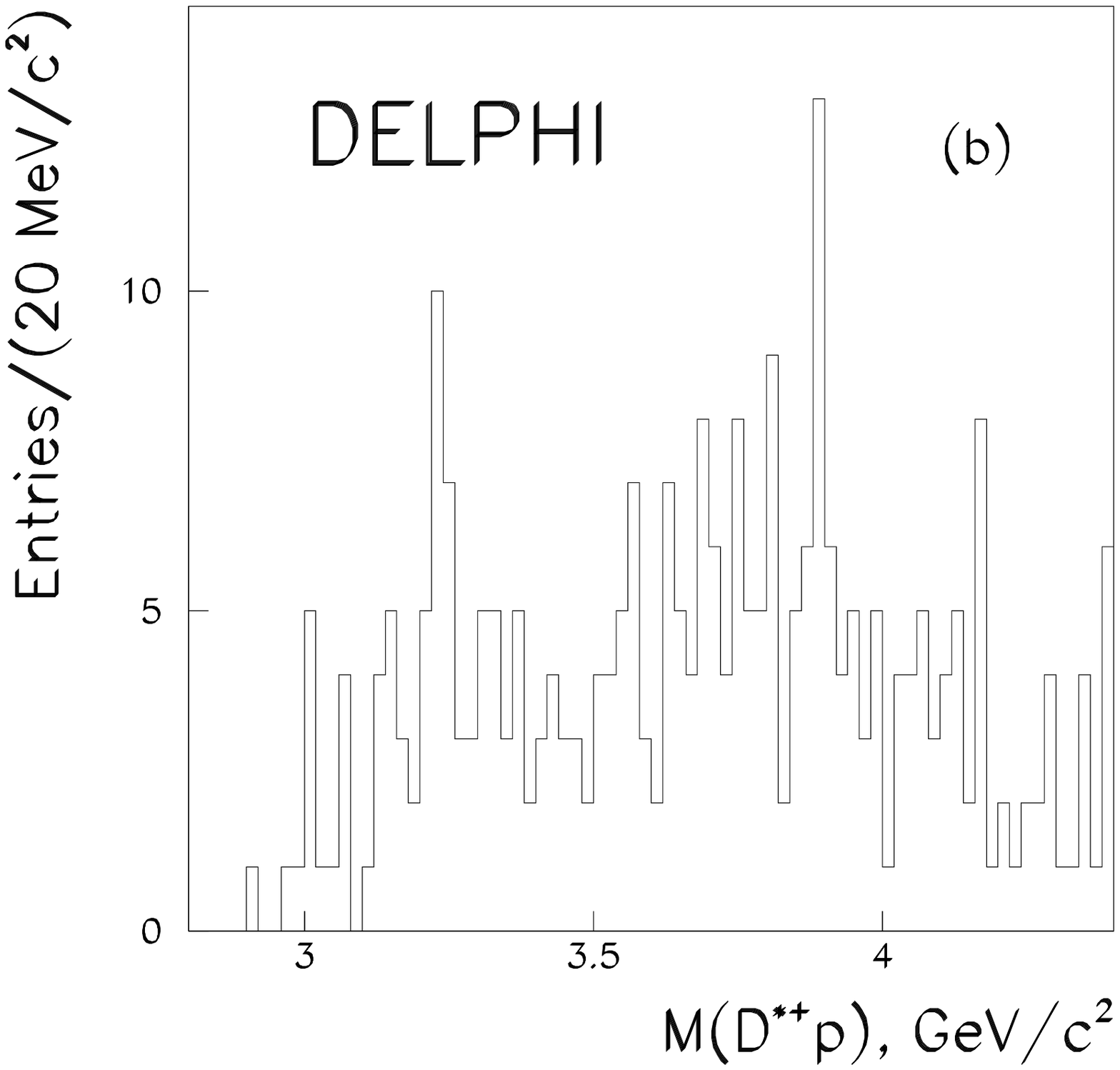,width=0.48\textwidth}}
\caption{\label{f:limthetac}Invariant masses (a) $M$(D$^{*+}\bar p$)  and  (b) $M$(D$^{*+}p$).}
\end{figure}

Figure \ref{f:limthetac} shows the invariant mass distributions of D$^*$p, for total charge
zero (right charge for a possible pentaquark) and total charge 2 (wrong charge) respectively.
No narrow resonance peak around 3.1 GeV/$c^2$
is seen in Figure \ref{f:limthetac} (a), which corresponds to the right charge.

To obtain an upper limit for the
production of a possible $\Theta_c(3100)^0$ state, a pentaquark signal was simulated; the detection efficiency 
for a $\Theta_c(3100)^0$ state decaying into D$^{*+}\bar p$ was estimated to be about 0.8\%, taking into account  the 
relevant branching fractions of the D$^*$ and of the D$^0$.

The best fit to the mass distribution of right charge pairs for a mass of 3100 MeV/$c^2$ and a width corresponding to the
experimental resolution, with the same procedure as described in the previous sections, gives an excess of $7 \pm 4$ events. The systematic uncertainties, dominated by the uncertainties on particle identification efficiencies, 
are negligible with respect to the error from the fit.
The 95\% CL upper limit on the average production rate, per hadronic Z decay, of
a $\Theta_c(3100)^0$ object
 decaying into D$^{*+}\bar p$, is 
\begin{equation} \langle N_{\Theta_c(3100)^0} \rangle \times Br(\Theta_c(3100)^0 \rightarrow D^{*+}\bar p)\; < 8.8 \times 10^{-4} \, .\end{equation}


\section{Conclusions}

A search for pentaquarks in hadronic Z decays was performed, and none of the states searched for was found.
Upper limits were established at 95\% CL on the average production rates
$\langle N \rangle$ of such particles and their charge-conjugate state per hadronic Z decay:
\begin{eqnarray*}
\langle N_{\Theta^{+}} \rangle \times Br(\Theta^+ \rightarrow pK^0_S) & < & 5.1 \times 10^{-4} \\
\langle N_{\Theta^{++}} \rangle & < & 1.6 \times 10^{-3} \\
\langle N_{\Phi(1860)^{--}} \rangle \times Br(\Phi(1860)^{--} \rightarrow \Xi^- \pi^-) & < & 2.9 \times 10^{-4} \\
\langle N_{\Theta_c(3100)^{0}} \rangle \times Br(\Theta_c(3100)^0 \rightarrow D^{*+}\bar p) & < & 8.8 \times 10^{-4}  \; .
\end{eqnarray*}
These limits improve previously published results ~\cite{aleph}.

In recent years thermodynamical \cite{beccatini} and
phenomenological models \cite{chliapnikov,pei} have appeared,
which successfully describe the overall particle production rates
in high energy interactions with very few parameters. According to
the model by Becattini \cite{beccatini}, the average production
rate for the production of the $\Theta^+$ at the Z energy should
be  of 0.007. According to the model by Chliapnikov
and Uvarov \cite{chliapnikov}, the average production rate is
expected to be less than 5 $\times 10^{-6}$, if the
$\Theta^+$ is dominantly produced from the intermediate
$N^*/\Delta^*$ baryon state with the mass of 2.4 GeV/$c^2$ as
indicated by the CLAS experiment \cite{pqsper}. On the other hand,
if the $\Theta^+$ production mechanism is similar to the one for
ordinary baryons produced at LEP, its average production rate
should be comparable with that of a known resonance,
the $\Lambda$(1520), which is observed with an average production
rate of $0.0224 \pm 0.0027$ per hadronic event \cite{pdg}.

\subsection*{Acknowledgements}
\vskip 3 mm

We thank Emile Schyns, Francesco Becattini and Rudi Fruewirth for
comments and suggestions.

We are greatly indebted to our technical 
collaborators, to the members of the CERN-SL Division for the excellent 
performance of the LEP collider, and to the funding agencies for their
support in building and operating the DELPHI detector.\\
We acknowledge in particular the support of \\
Austrian Federal Ministry of Education, Science and Culture,
GZ 616.364/2-III/2a/98, \\
FNRS--FWO, Flanders Institute to encourage scientific and technological 
research in the industry (IWT) and Belgian Federal Office for Scientific, 
Technical and Cultural affairs (OSTC), Belgium, \\
FINEP, CNPq, CAPES, FUJB and FAPERJ, Brazil, \\
Ministry of Education of the Czech Republic, project LC527, \\
Academy of Sciences of the Czech Republic, project AV0Z10100502, \\
Commission of the European Communities (DG XII), \\
Direction des Sciences de la Mati$\grave{\mbox{\rm e}}$re, CEA, France, \\
Bundesministerium f$\ddot{\mbox{\rm u}}$r Bildung, Wissenschaft, Forschung 
und Technologie, Germany,\\
General Secretariat for Research and Technology, Greece, \\
National Science Foundation (NWO) and Foundation for Research on Matter (FOM),
The Netherlands, \\
Norwegian Research Council,  \\
State Committee for Scientific Research, Poland, SPUB-M/CERN/PO3/DZ296/2000,
SPUB-M/CERN/PO3/DZ297/2000, 2P03B 104 19 and 2P03B 69 23(2002-2004), \\
FCT - Funda\c{c}\~ao para a Ci\^encia e Tecnologia, Portugal, \\
Vedecka grantova agentura MS SR, Slovakia, Nr. 95/5195/134, \\
Ministry of Science and Technology of the Republic of Slovenia, \\
CICYT, Spain, AEN99-0950 and AEN99-0761,  \\
The Swedish Research Council,      \\
Particle Physics and Astronomy Research Council, UK, \\
Department of Energy, USA, DE-FG02-01ER41155, \\
EEC RTN contract HPRN-CT-00292-2002. \\



\end{document}